\documentclass{aa}  

\usepackage{txfonts}
\usepackage{hyperref}
\usepackage[flushleft]{threeparttable}
\usepackage{graphicx} 
\usepackage{subcaption} 

\usepackage[dvipsnames]{xcolor} 
\usepackage{color}
\definecolor{mhi}{rgb}{0.6,0.0,0.6}

\defcitealias{Buzzo_25c}{Paper~I}
\usepackage{multirow}
\begin{document}

\title{GAMA~526784: the progenitor of a globular cluster-rich ultra-diffuse galaxy?}

\subtitle{II. Molecular gas, neutral gas and environment}

\author{Maria Luisa Buzzo\inst{1,2,*}
\and Anita Zanella\inst{3}
\and Michael Hilker\inst{1}
\and Kristine Spekkens\inst{4}
\and Laura Hunter\inst{5}
\and Laurella C. Marin\inst{5}
}

\institute{European Southern Observatory, Karl-Schwarzschild-Strasse 2, 85748 Garching bei M\"unchen, Germany 
\and Centre for Astrophysics and Supercomputing, Swinburne University, John Street, Hawthorn VIC 3122, Australia 
\and INAF - Osservatorio di Astrofisica e Scienza dello Spazio di Bologna, Via Gobetti 93/3, I-40129, Bologna, Italy
\and Department of Physics, Engineering Physics, and Astronomy, Queen’s University, Kingston, ON, K7L 3N6, Canada
\and Department of Physics and Astronomy, Dartmouth College, Hanover, NH 03755, USA
\\ *\email{luisa.buzzo@gmail.com}
         }

   \date{Received XX; accepted YY}
 
  \abstract
  {}
  {We investigate the gas reservoirs, star formation properties, and environment of the ultra-diffuse galaxy GAMA~526784 to understand its formation history, the efficiency of molecular gas conversion into stars, and the possible role of an interacting companion in shaping its current morphology.}
{We analyse low- and high-resolution CO observations to place constraints on the molecular gas content of the galaxy, compare them with HI data, and examine the star formation efficiency of GAMA~526784. The potential influence of a newly identified nearby dwarf galaxy is assessed using photometric and spatial information.}
{GAMA~526784 exhibits a regular HI reservoir ($M_{HI}/M_{\star} = 2.88$) but only upper limits on its molecular gas mass ($M_{H_2}(5\sigma)/M_{\star}  < 0.23$). The galaxy's HI reservoir and CO non-detection can be explained by several mechanisms: (1) the predominance of CO-dark H$_2$, which remains invisible to CO observations but contributes to star formation; (2) a time delay in HI-to-H$_2$ conversion following a recent interaction; or (3) elevated turbulence inhibiting gas collapse. An identified companion, optically found at a projected distance of $\sim48$ kpc, shows similar colours and lies in the direction of the young star clusters in GAMA~526784, indicating a possible association. We hypothesise that this companion may have triggered the formation of the star clusters in GAMA~526784 through a high-velocity encounter.}
{Our findings suggest that GAMA~526784 may have undergone a dwarf-dwarf interaction that significantly influenced its gas reservoirs and star formation activity. The presence of a nearby companion galaxy is consistent with predictions of a high-speed encounter, potentially offering a rare observational example of such an interaction in progress. We hypothesise that this encounter may have played a key role in shaping the system's recent evolution. Future observations, particularly targeting molecular gas tracers beyond CO and resolved HI observations, will be crucial in determining the true extent of GAMA~526784’s cold gas reservoir and the nature of its recent star formation activity.}

   \maketitle
%

\section{Introduction}
\label{sec:introduction}

Ultra-diffuse galaxies (UDGs) continue to challenge our understanding of galaxy formation and evolution. Defined by their low surface brightness ($\mu_{g,0}>24$ mag arcsec$^{-2}$) and large effective radii ($R_{\rm e}>1.5$ kpc), UDGs are found in diverse environments, from dense galaxy clusters to the low-density field. Their origins remain debated, with two primary formation pathways emerging: the "failed galaxy" scenario, where early quenching prevents the typically extended star formation \citep{vanDokkum_15, Danieli_22}, and the "puffed-up dwarf" scenario, in which low-mass galaxies are expanded by internal and external processes \citep{Amorisco_16,DiCintio_17, Carleton_19}. However, recent discoveries suggest that some UDGs may form through alternative mechanisms, including high-speed interactions or collisions \citep{Silk_19, vanDokkum_22,Shin_20,Lee_21,Lee_24}.

One such case is GAMA~526784, an isolated UDG exhibiting a striking two-component structure: a compact, quiescent inner stellar body and an extended, actively star-forming outer region \citep[][hereafter \citetalias{Buzzo_25c}]{Buzzo_25c}. In \citetalias{Buzzo_25c}, we analysed deep imaging and integral field spectroscopy of the system, revealing a mass-weighted stellar age of $\sim$9.9~Gyr and metallicity of [M/H]~$\sim -1.0$ in the central body, and a significantly younger ($\sim$0.9~Gyr), more metal-poor ([M/H]~$\sim -1.2$) population in the outskirts. Star formation is ongoing and highly localised, with a total SFR of $\sim$0.01~M$_\odot$~yr$^{-1}$ concentrated in compact clumps aligned in a $\sim$5~kpc string-like structure. These regions show elevated electron densities (inferred from [SII]$\lambda$6716/$\lambda$6731~$\sim 1.1$), significant dust attenuation ($A_V$ up to 2.2~mag), and gas velocity dispersions reaching 50~km~s$^{-1}$, all indicative of dense, possibly shock-compressed star-forming regions. The ionised gas is kinematically misaligned from the stellar body by $\sim$20$^\circ$, supporting the idea of a recent external perturbation. In total, 29 star cluster candidates were identified, including a population of young clusters with ages of 8--11~Myr and stellar masses of $\log(M_\star/\mathrm{M}_\odot)\sim5.0$, and a more metal-poor, old GC population with ages of $\sim$9~Gyr and masses of $\log(M_\star/\mathrm{M}_\odot)\sim5.5$. This bimodal star cluster population, along with its gas and stellar properties, raises the possibility that GAMA~526784 could represent a transitional phase in the formation of GC-rich UDGs.

While the optical data provide strong evidence for a recent interaction, the nature of this event remains unclear. The "bullet dwarf" scenario, in which a high-speed collision separates baryonic and dark matter and triggers a burst of star cluster formation \citep{Silk_19, vanDokkum_22}, remains a compelling possibility. However, alternative explanations, such as a minor merger with a gas-rich dwarf, could also account for the observed properties \citep{Fensch_19b}. To distinguish between these scenarios, a detailed analysis of the galaxy’s molecular and neutral gas content, as well as its large-scale environment, is essential.

This paper presents Atacama Large Milimeter/submilimeter Array (ALMA) and Green Bank Telescope (GBT) observations of GAMA~526784 to probe its molecular and neutral gas reservoirs. These data will allow us to assess the availability of gas for future star formation, determine whether the observed shocks are associated with a gas-rich merger, and evaluate whether the system has sufficient gas to sustain massive star cluster formation. If GAMA~526784 is indeed forming young star clusters that will evolve into GCs, some of these young clusters (with ages $< 3-5$ Myr) might still be embedded in giant molecular clouds \citep[GMCs,][]{Lada_03,Kim_21}. If the clusters have available gas to become more massive, this would provide direct support for the galaxy becoming GC-rich and offer new insights into how UDGs evolve over time.

By combining ionised, molecular, and neutral gas data, this study aims to clarify the evolutionary pathway of GAMA~526784 and its implications for UDG formation models. More broadly, understanding the gas content and star formation potential of this system contributes to the ongoing effort to map the diversity of UDG evolutionary pathways and to determine the role of high-speed interactions in shaping these enigmatic galaxies.

This paper is structured as follows: in Section 2, we describe the ALMA and GBT observations and data reduction processes. In Section 3, we present the methodology used to derive molecular and neutral gas measurements. Section 4 discusses the results and their implications for the formation of GAMA~526784, and we compare our findings to other UDGs and galaxy types to place GAMA~526784 in context. Finally, Section 5 summarises our conclusions.

\section{Data}
\label{sec:data}

To investigate the properties of GAMA~526784, we compiled a comprehensive multiwavelength dataset. Our analysis builds upon previous optical observations (\citetalias{Buzzo_25c}), which provided insights into the star clusters, stellar populations, and ionised gas properties of GAMA~526784. In this work, we extend our study to the cold gas component by utilizing observations from ALMA and GBT. The ALMA data allow us to trace the molecular gas through CO($1-0$) emission, probing the star-forming potential of the system, while the GBT data provide a measurement of its neutral hydrogen (HI) reservoir, essential for understanding the galaxy's overall gas content and potential for continued star formation. Below, we describe the details of these observations and data reduction procedures.

\subsection{ALMA Observations}
\label{sec:alma_data}

We carried out the first ever ALMA observations of an ultra-diffuse galaxy using ALMA's Band 3 during Cycle 8 (PI: A. Zanella, Project ID: 2021.1.00898.S) with the goal of detecting the CO($1-0$) emission line at a rest-frame frequency of $\nu_{\rm rf} = 115.271202$ GHz and the underlying continuum, redshifted to an observed frequency of $\nu_{\rm obs} = 114.325475$ GHz, assuming the observed radial velocity of the galaxy of $V = 2758$ km s$^{-1}$ (\citetalias{Buzzo_25c}). The diffuse body of GAMA~526784 (hereafter `low-resolution') was observed for 3 hours with the 7m array, while the higher-resolution dataset focused on the star clusters and their parent GMCs (hereafter `high-resolution') was observed for nearly 6.5 hours with the 12m array, reaching sensitivities of 1.0~mJy~beam$^{-1}$ and 9~$\mu$Jy~beam$^{-1}$, respectively. 

Observations were conducted between November and December 2021 for the low-resolution data and between November 2021 and July 2022 for the high-resolution data.

The native spectral resolutions of the observations are 10.255 km s$^{-1}$ for the low-resolution data and 1.282 km s$^{-1}$ for the high-resolution data. The imaged beam sizes for the low-resolution data are FWHM = $15.8'' \times 9.9''$, and the observations were carried out with the Atacama Compact Array (ACA). For the high-resolution data, the beam sizes are FWHM = $0.74'' \times 0.56''$. Table \ref{tab:alma} summarises the ALMA observations.

\begin{table*}
\centering
\caption{Summary of ALMA and GBT observations of GAMA~526784, including array configuration, exposure time, beam size, spectral resolution, and sensitivity.}
\begin{tabular}{lccc}
& \multicolumn{2}{c}{ALMA} & GBT \\ \hline
& Low-resolution & High-resolution &  -- \\ \hline
Array & 7M & 12M & -- \\
Exposure time (hours) & 3.01 & 6.45 & 2.67 \\
Beam sizes & $15.8'' \times 9.9''$ & $0.74'' \times 0.56''$ & $9' \times 9'$\\
Spectral Resolution (km s$^{-1}$) & 10.255 & 1.282 & 0.15 \\ 
Sensitivity (mJy~beam$^{-1}$) & 1.0 & 0.009 & 2.6 \\ \hline
\end{tabular}
\label{tab:alma}
\end{table*}

The data were reduced using the standard ALMA pipeline with the \texttt{CASA} software \citep{McMullin_07}. The calibrated data cubes were converted to uvfits format and analysed in the uv-plane using \texttt{GILDAS} \citep{Guilloteau_Lucas_00}, which is typical for faint sources. The analysis in the uv-plane offers several advantages over the image plane: it uses raw interferometric data (visibilities) and avoids image reconstruction and deconvolution, eliminating biases and artefacts from cleaning and providing a model-independent view. It also accounts for the actual spatial frequencies sampled by the array, respecting the instrument’s resolution and sensitivity limits, which is particularly important for low-SNR data like ours. Additionally, it allows fine control over how different baselines are weighted or tapered helping isolate spatial scales or emphasise specific structures. For compact, unresolved, low-SNR sources such as ours, uv-plane analysis is therefore preferable, while image-plane analysis remains valuable for spatially visualising complex or extended sources.

\subsection{Green Bank Telescope Observations}
\label{sec:gbt_data}

GBT observations of GAMA~526784 were conducted on September 29, 2024, under project ID GBT24B-319. The total integration time was 160 minutes, achieved through a sequence of 600-second ON/OFF pairs (300~s ON, 300~s OFF).

We utilised the L-band receiver coupled with the Versatile GBT Astronomical Spectrometer (VEGAS) operating in Mode 10, providing a bandwidth of 23.44~MHz and a spectral resolution of 0.7~kHz. The spectral setup was centred at the HI 21cm line ($1.42041$ GHz) redshifted to correspond to a velocity of 2743 km s$^{-1}$ for GAMA~526784.

The L-band receiver provides a beam size of approximately 9~arcminutes and a native velocity resolution of 0.15~km\,s$^{-1}$. The data were reduced using the standard GBTIDL routine \texttt{getps}, followed by baseline subtraction and spectral smoothing to enhance the signal-to-noise ratio.

\section{Analysis}
\label{sec:analysis}
In what follows, we analyse the ALMA and GBT observations of GAMA~526784 to constrain its molecular and neutral gas content, assess its star formation efficiency, and investigate the role of its environment.

\begin{figure*}
    \centering
    \includegraphics[width=\textwidth]{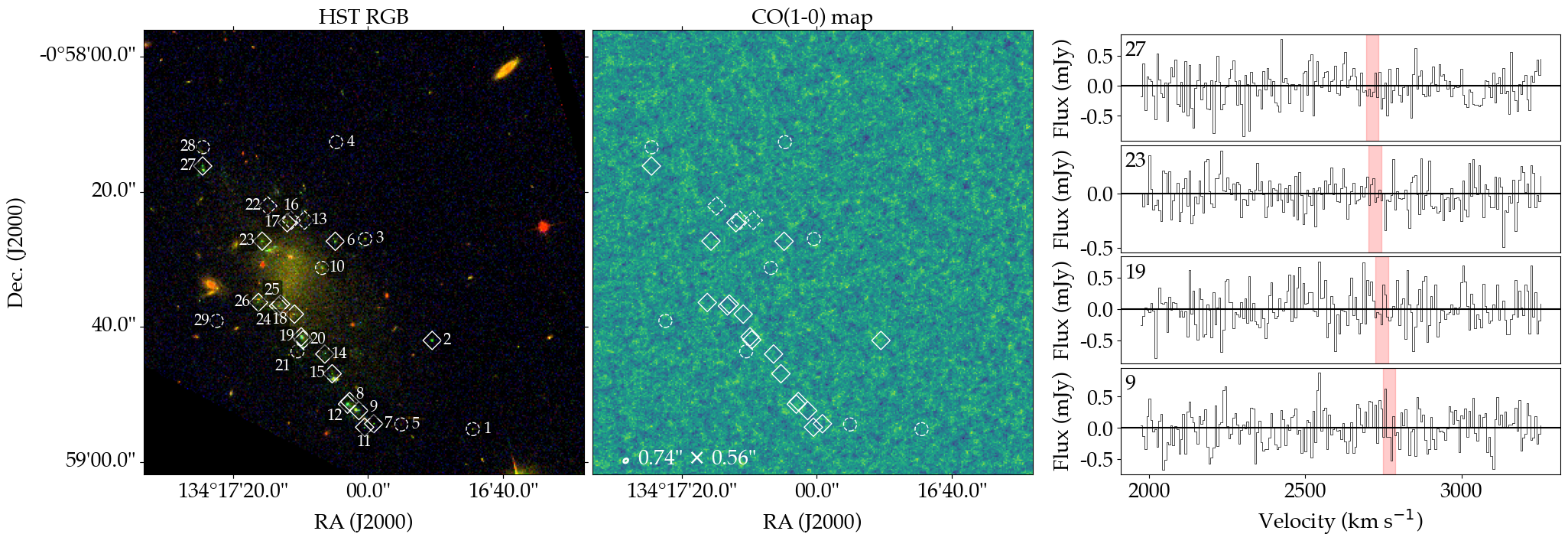}
    \caption{Comparison of the HST optical image with the ALMA high-resolution cube of GAMA~526784. The star clusters (which are expected to be embedded in GMCs) are marked with labels corresponding to their numbers, following the naming convention used in \citetalias{Buzzo_25c} The middle panel shows the high-resolution ALMA cube. The beam size and orientation are indicated by the white ellipse in the bottom-left corner. The spectra of the four brightest clusters are shown in the rightmost panel, with the expected CO($1-0$) emission wavelength marked by a red shade. No detectable CO($1-0$) emission is observed in any of the spectra.}
    \label{fig:maps}
\end{figure*}

\subsection{CO($1-0$) Emission}
To investigate the molecular gas content of GAMA~526784, we first constructed the velocity-integrated CO(1-0) line maps. We adopted the method used by \cite{Zanella_23} and \cite{Zanella_18}. The first step is to determine the optimal spectral range for integration. Since the measurement is performed directly in the uv-plane, we extracted the CO spectrum by applying the Fourier Transform of a PSF model and looked for a detection around the expected frequency of the CO emission based on the redshift of the galaxy. We adopt a PSF model because at the resolution of the ACA observations ($15.8'' \times 9.9''$), the galaxy is unresolved with an effective radius of 13.2 arcsec (\citetalias{Buzzo_25c}).

For the PSF modeling, we fixed the central coordinates based on the optical images (see \citetalias{Buzzo_25c}), allowing the flux to vary across the channels. We then analysed the resulting spectrum to detect any positive emission line signals. However, no CO emission was detected in the spectra, suggesting that the diffuse body of GAMA~526784 does not host a significant amount of molecular gas that is traced by CO($1-0$) emission.  

We then averaged the data over the channels where the CO line was expected, based on the galaxy's redshift as determined by \citetalias{Buzzo_25c} (i.e. 2758 km s$^{-1}$). Since the CO(1-0) line is at $\nu = 115.271202$ GHz at restframe, applying the redshift of the galaxy, we expected the CO(1-0) line at a frequency of $\nu = 114.2316936$ GHz. We found the channel corresponding to this frequency and averaged the neighbouring channels allowing for a velocity width of 30 km s$^{-1}$ up and down, i.e. three channels at each side. No CO($1-0$) signal was detected in the averaged map of the low-resolution data targetting the diffuse body of the galaxy. Although this whole analysis was done in the uv-plane, we also inspected the imaging and arrived at the same conclusions, i.e. no CO detection. The final averaged CO map of the low-resolution data is shown in Appendix \ref{sec:appendix_diffuse}.

We applied the same method to the high-resolution data cube, using the star cluster positions identified by \citetalias{Buzzo_25c}, assuming that star clusters and their parent GMCs are co-spatial or are offset by less than the beam size ($0.74'' \times 0.56''$) and that GMCs are unresolved at the resolution of our observations and fitting them as point sources. For the average map, we assumed a velocity width of 10 km s$^{-1}$ around the expected CO line. Again, no CO($1-0$) emission was detected at the location of any star clusters. All data has been corrected for primary beam attenuation. 

The resulting averaged map and spectra extracted at the location of the brightest star clusters in GAMA~526784 are presented in Figure \ref{fig:maps}. The spectra of the fainter clusters is of similar quality and also do not show any clear sign of CO emission.

Although no CO emission was detected, we estimate upper limits on the molecular gas mass using the $5\sigma$ uncertainties in the flux measurements derived from the point-source fits.

\subsection{Continuum Emission}
In a similar approach to the CO emission, we generated continuum maps by integrating over the entire spectral range, excluding the channels where CO($1-0$) emission was expected based on the galaxy’s redshift.  The continuum was not detected neither in the low- or the high-resolution data, indicating a lack of significant emission in this spectral range as well.

\subsection{Molecular Gas Mass}
To estimate molecular gas mass constraints from the upper limits of our CO flux densities, we primarily adopt a metallicity-dependent CO-to-H$_2$ conversion factor, which is more appropriate for the low-metallicity, low-mass environment of GAMA~526784 \citep{Accurso_17, Ramambason_24, Cormier_14, Shi_14}. For comparison, we also calculate masses using the standard Milky Way conversion factor, which is a commonly used benchmark in extragalactic studies.

The metallicity-dependent conversion, introduced by \citet{Accurso_17}, accounts for both the gas-phase metallicity and the galaxy’s offset from the star formation main sequence (SFR--$M_\star$ relation) of \citet{Whitaker_12}, and is given by:

\begin{equation}
\begin{aligned}
    \log \alpha_{\rm CO} (\pm 0.165\, {\rm dex}) &= 15.623 - 1.732\,[12 + \log(\rm{O/H})] \\
    &+ 0.051 \,\log \Delta(\rm MS),
\end{aligned}
\end{equation}

where $12 + \log({\rm O/H})$ represents the gas-phase metallicity, and $\Delta(\rm{MS})$ is the distance of the galaxy from the star formation main sequence, defined as:

\begin{equation}
\Delta (\rm MS) = \frac{\rm sSFR_{measured}}{\rm sSFR_{ms}(z,M_{\star})},
\end{equation}

where ${\rm sSFR}_{\rm measured}$ is the measured specific star formation rate (sSFR), and ${\rm sSFR_{ms}}$ is the star formation rate predicted for a galaxy on the star formation main sequence. The latter is given by:

\begin{equation}
    \begin{aligned}
        \log (\rm sSFR_{ms} (z, M_{\star})) &= -1.12 + 1.14z - 0.19z^2 - (0.3 + 0.13z) \\
        &\times (\log M_{\star} - 10.5) \rm{[Gyr^{-1}]},
    \end{aligned}
\end{equation}

where $z$ is the galaxy redshift, and $M_{\star}$ is its stellar mass. Using this approach, the velocity-integrated CO line luminosity can be converted into molecular gas mass using the standard methods thoroughly explained in \cite{Solomon_VandenBout_05}. Assuming a global gas-phase metallicity of $12 + \log(\mathrm{O}/\mathrm{H}) = 8.3$ \citep[\citetalias{Buzzo_25c}; derived using the O3N2 calibration from][]{Marino_13}, and the recovered $\alpha_{\rm CO,met} = 17.01 M_\odot\, (\mathrm{K\, km\, s^{-1}\, pc^2})^{-1}$ to derive 5$\sigma$ upper limits on the molecular gas content of GAMA~526784, which leads to $\log(M_{\rm H_2}/M_{\odot}) \lesssim 7.6 \, M_{\odot}$.

For comparison, we also compute the molecular gas mass using the traditional Milky Way conversion factor, $\alpha_{\rm CO, MW} = 4.36\, M_\odot\, (\mathrm{K\, km\, s^{-1}\, pc^2})^{-1}$ \citep{Strong_Mattox_96, Abdo_10}, which is frequently applied to galaxies with solar-like metallicities and has been used in studies such as the PHANGS survey \citep[e.g.,][]{Leroy_21}. This yields a molecular gas mass estimate approximately four times smaller than those obtained using the metallicity-dependent method, highlighting the importance of applying conversion factors tailored to low-metallicity systems \citep[e.g.,][]{Accurso_17, Ramambason_24, Cormier_14, Shi_14}.

We extend the same metallicity-based method to individual GMCs, using the star formation rates and metallicities measured within the apertures defined for their flux extraction. While CO-to-H$_2$ conversion in solar-metallicity GMCs often assume the Milky Way value \citep{Bolatto_13}, systems like GAMA~526784 ([M/H]~$\simeq -1.5$ dex) are expected to require significantly larger $\alpha_{\rm CO}$ values due to reduced dust shielding and enhanced photodissociation of CO \citep{Schruba_12}. For comparison, resolved studies of GMCs in the LMC ([M/H]~$\sim -0.3$ dex) yield $\alpha_{\rm CO} = 6.9 \pm 2.3$ \citep{Hughes_10}, and the SMC ([M/H]~$\sim -0.7$ dex) shows conversion factors exceeding $\alpha_{\rm CO} > 50$ \citep{Bolatto_08}. In our case, the metallicity-dependent conversion results in $\alpha_{\rm CO}$ values ranging from 6 to 25 across different GMCs, consistent with expectations for both more metal-rich and metal-poor cloud populations. The recovered physical properties of GAMA~526784 and the GMCs are shown in Table~\ref{tab:sfr_metallicity}.

The resulting upper limits on individual GMC masses ($M_{\rm H_2} < 10^{6.0} - 10^{7.0}\, M_\odot$) fall within the range of the lowest-mass GMCs detected in PHANGS galaxies \citep{Sun_20}. These values are also comparable to those of well-known local clouds, including the most massive Milky Way GMC ($6 \times 10^6\, M_\odot$), the largest M33 GMC ($7 \times 10^5\, M_\odot$), and the Orion complex ($5 \times 10^5\, M_\odot$) \citep{Engargiola_03, Wilson_05}.

\subsection{HI Mass Calculation}
\begin{figure}
    \centering
    \includegraphics[width=\columnwidth]{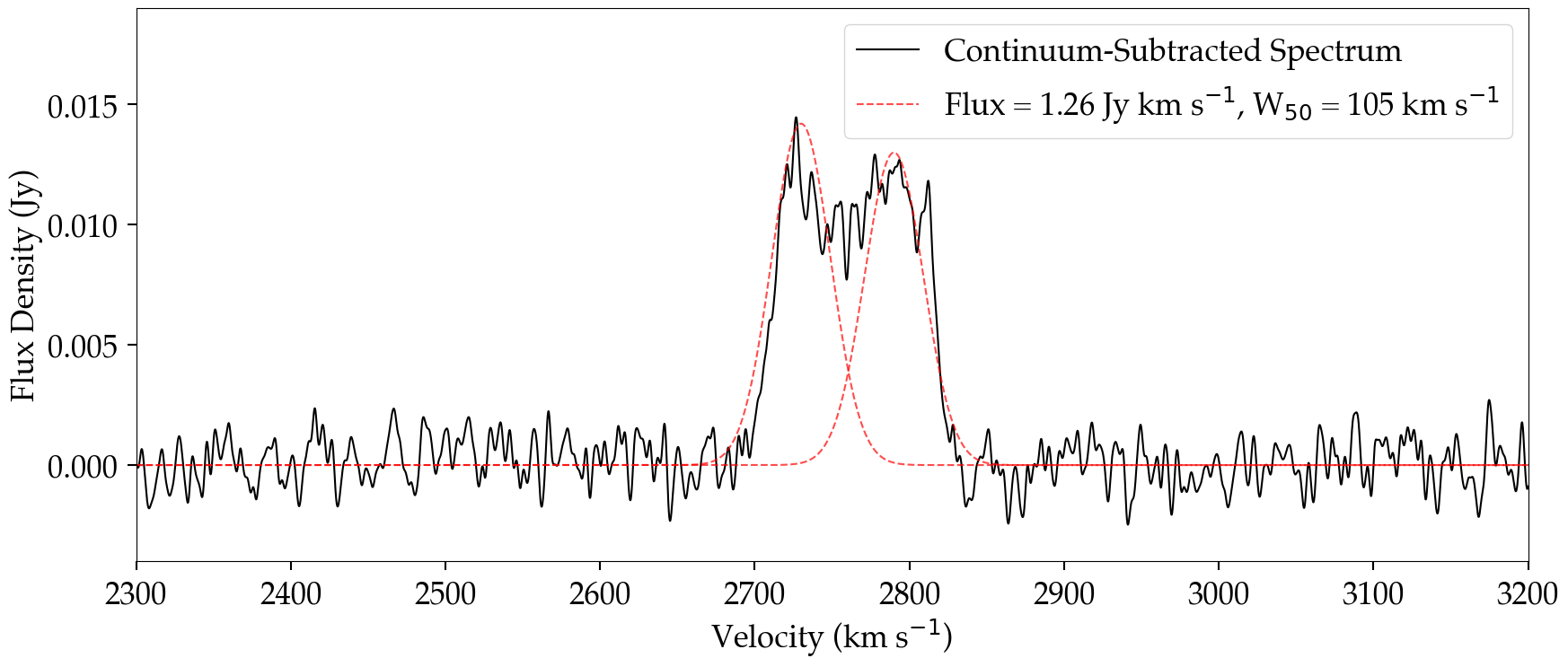}
    \caption{HI spectrum of GAMA~526784, smoothed by a factor of 3 (resulting in a spectral resolution of 0.45~km\,s$^{-1}$). The profile exhibits a clear double-horned shape, characteristic of a rotating disk, with a velocity width at 50\% of the peak flux ($W_{50}$) of 105~km\,s$^{-1}$. Integrating the emission yields a total HI flux of 1.26~Jy\,km\,s$^{-1}$.}
    \label{fig:Hi_spectrum}
\end{figure}

In addition to the molecular gas, we estimate the HI mass of GAMA~526784 using the GBT data. The HI spectrum (Figure~\ref{fig:Hi_spectrum}) shows a double-horned profile characteristic of a rotating disk centred at V = 2758 km s$^{-1}$ with a velocity width of the HI line profile, at 50\% of the peak intensity, of $W_{50}=105$ km s$^{-1}$. The integrated HI flux density is $S_{\rm HI}=1.26\pm0.13$ Jy km s$^{-1}$ at the native resolution of 0.15 km s$^{-1}$.

The HI mass was calculated using the standard relation:
\begin{equation}
M_{\rm HI} = 2.36 \times 10^5 \times D_L^2 \times S_{\rm HI}\, M_{\odot},
\label{eq:hi}
\end{equation}
where $D_L=40$ Mpc is the luminosity distance. This yields $\log(M_{\rm HI}/M_{\odot}) = 8.7 \pm 0.9$.
With the recovered $W_{50}$, one can estimate the rotation velocity of the galaxy and its dynamical mass, using the approach of \cite{Spekkens_18}. We assume that $W_{50} = 2\,V_{\rm rot} \sin\, i$, where $\sin\,~i~=~\sqrt{\frac{1-(b/a)^2}{1 - q_0^2}}$, $q_0 = 0.2$ \citep{Hubble_26, Haynes_84,Bradford_16} and $b/a = 0.52$ (\citetalias{Buzzo_25c}). This renders a rotation velocity for GAMA~526784 of $V_{\rm rot} = 59.3 \pm 5.9$ km s$^{-1}$, after correcting for inclination. This value is consistent with the ionised gas rotation amplitude (\citetalias{Buzzo_25c}). To calculate the dynamical mass of the galaxy, we assume that the HI gas distribution extends out to three effective radii \citep{Broelis_Rhee_97}, and use:

\begin{equation}
    M_{\rm dyn}^{3Re} = 3.5 \times 10^5 R_e W_{50}^2 M_{\odot},
\end{equation}

with $R_e = 2.5$ kpc (\citetalias{Buzzo_25c}). We estimate a dynamical mass within 3$R_e$ of $\log(M_{\rm dyn}^{3R_e}/M_{\odot}) = 10.0 \pm 1.0$. The baryonic mass, derived from the stellar mass ($\log(M_{\star}/M_{\odot}) \sim 8.3$) and HI mass ($\log(M_{\rm HI}/M_{\odot}) \sim 8.7$), totals approximately $\log(M_{\rm baryonic}/M_{\odot}) \sim 8.9$. This implies that the galaxy is likely dark matter-dominated within $3~R_e$.

\begin{figure}
    \centering
    \includegraphics[width=0.9\columnwidth]{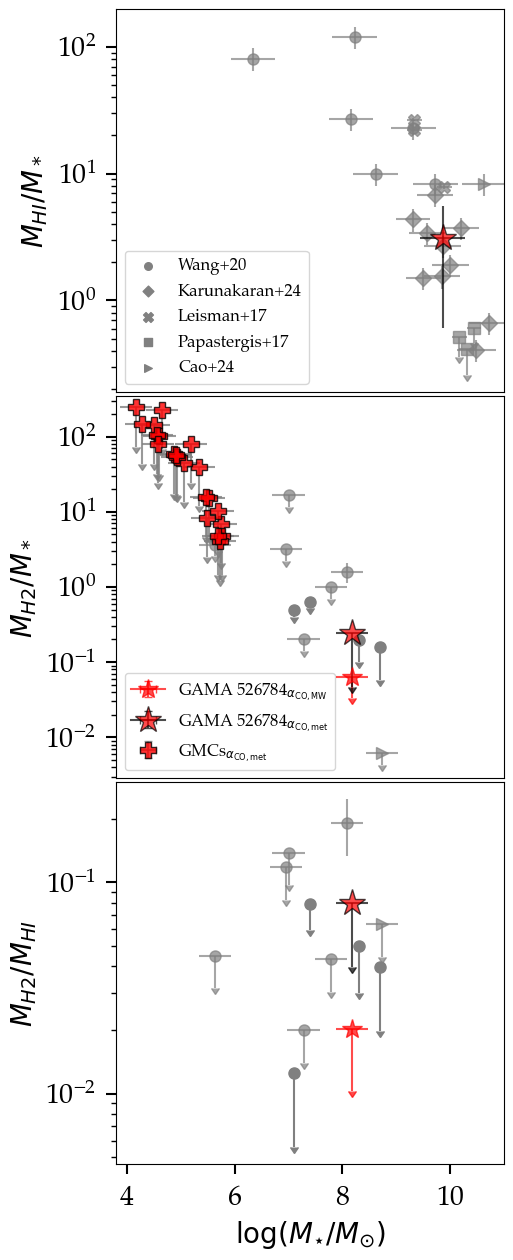}
    \caption{Scaling relations for the molecular and neutral has content of disc-dominated galaxies as a function of stellar mass. Grey scatter points are literature values from \cite{Karunakaran_24}, \cite{Leisman_17}, \cite{Papastergis_17}, \cite{Wang_20} and \cite{Cao_24}. Points with arrows pointing downwards are 5$\sigma$ upper limits. GAMA 526784 is shown in all panels with the red star symbol. The black edgecolor marks measurement using a metallicity dependent CO-to-$H_2$ conversion \citep{Accurso_17}. The one without any edgecolor uses the standard MW $\alpha_{\rm CO}$ conversion. GMCs are shown with plus signs, with their masses calculated using also the metallicity dependent conversion.}
    \label{fig:HI_H2_comparison}
\end{figure}

\section{Results and Discussion}
\label{sec:discussion}

GAMA~526784 provides a valuable case study of gas-rich, low-mass galaxies, with its significant neutral hydrogen reservoir and the non-detection of molecular gas traced by CO($1-0$) emission. Figure~\ref{fig:HI_H2_comparison} compares GAMA~526784 with literature data for HI and H$_2$ masses as a function of stellar mass. The top panel shows that the HI properties of GAMA~526784 are typical for a star-forming galaxy of its stellar mass. Its $M_{\rm HI}/M_*$ ratio of $\sim$2.9 is consistent with other gas-rich UDGs and LSBs from \citet{Karunakaran_24}. In contrast, quiescent field UDGs like those studied by \citet{Papastergis_17} ($M_{\rm HI}/M_*\sim0.5$) have a much lower $M_{HI}/M_{\star}$ ratio, likely due to environmental gas stripping or suppressed star formation. The galaxy's HI content is also less extreme than that of HI-selected systems like those in \citet{Leisman_17}, which are biased towards higher gas fractions due to their selection criteria.

The middle panel of Figure~\ref{fig:HI_H2_comparison} depicts the molecular-to-stellar mass ratio, highlighting the stringent upper limits on the CO-derived H$_2$ mass.
Recent studies of CO in UDGs and LSBs are shown for comparison in the figure (e.g., \citealt{Wang_20, Cao_24}). These works have found similarly low CO-detected H$_2$ masses. It is important to note that for a correct comparison between the studies of \cite{Wang_20}, \cite{Cao_24} and ours, we convert their 3$\sigma$ upper limits in the $H_2$ mass estimates into 5$\sigma$ upper limits assuming Gaussian statistics.

The galaxy's HI reservoir ($M_{\rm HI} = (4.7\pm0.5)\times10^8$ M$_\odot$) and CO non-detection can be explained by several mechanisms: (1) the predominance of CO-dark H$_2$, which remains invisible to CO observations but contributes to star formation; (2) a time delay in HI-to-H$_2$ conversion following a recent interaction \citep{Cullen_07, Renaud_19}; or (3) elevated turbulence inhibiting gas collapse \citep{Krumholz_12}. The non-detection of CO despite active star formation is consistent with trends observed in other low-metallicity dwarfs \citep{Leroy_08,Cormier_14, Ramambason_24}, reinforcing that GAMA~526784 is not an outlier but rather a representative example of its class.

\begin{figure*}[t]
    \centering
    \includegraphics[width=0.9\textwidth]{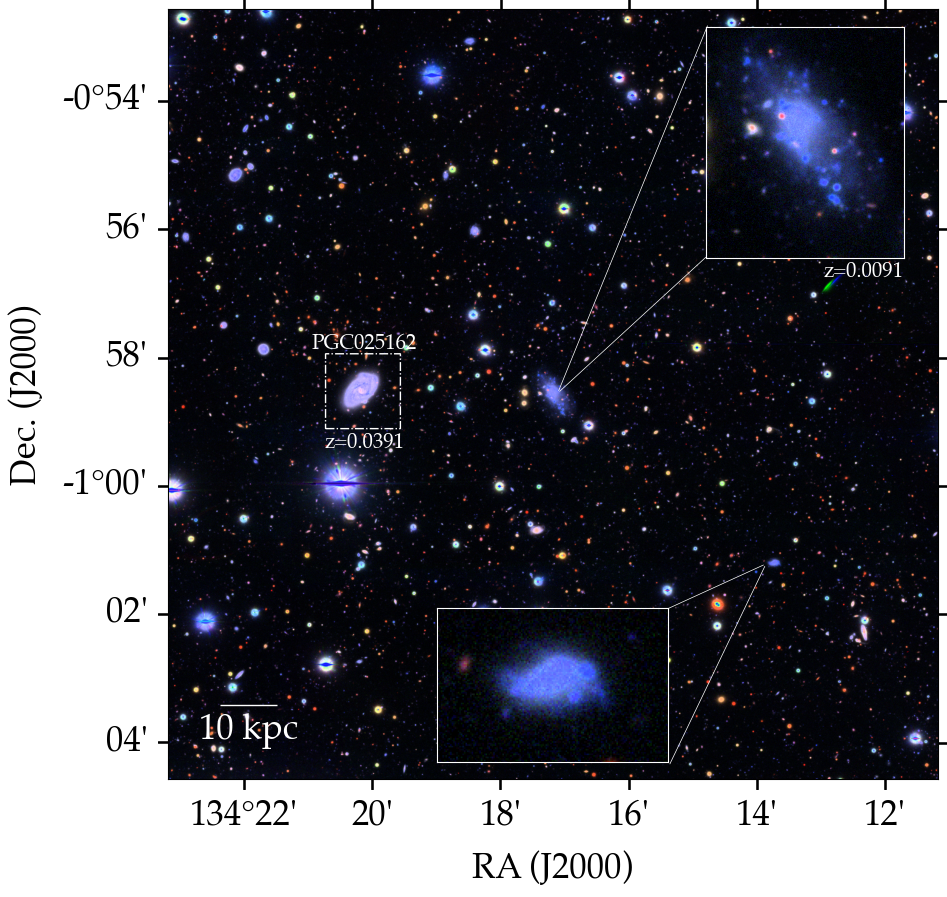}
    \caption{The possible interacting system of GAMA~526784 and putative companion, separated by 48 kpc in projection. Young star clusters in GAMA 526784 align along the axis possibly connecting the galaxies, with their ages and the star formation history of the galaxy suggesting formation during a close encounter $\sim$100 Myr ago. This spatial and possibly kinematic configuration provides compelling evidence for a physical association between the galaxies. The image is 12 arcmin on the side. North is up and east is to the left. The only massive galaxy close in projection to GAMA~526784 is PGC~025165 (highlighted in the dotted rectangle) and it lies at a redshift of $z=0.0391$, i.e. far in the background.}
    \label{fig:companion}
\end{figure*}

\subsection{Environment}
GAMA~526784 is an isolated field UDG, with no obvious massive companion galaxies in its vicinity. In projection, the closest galaxy is PGC~025162 (highlighted in Figure~\ref{fig:companion}), located approximately 3 arcmin away ($\sim$34 kpc at the redshift of GAMA~526784). However, its redshift from the HyperLEDA survey ($z = 0.0391$) places it well in the background compared to GAMA~526784 ($z = 0.0091$), confirming that the two are not physically associated. A comprehensive search of the surrounding region using photometric and spectroscopic redshifts from the GAMA catalogue reveals no massive galaxies within several hundred kiloparsecs. A larger-scale view of the environment, presented in Figure~\ref{fig:companion}, spans a $12 \times 12$~arcmin$^2$ field-of-view and clearly illustrates the lack of nearby companions, reinforcing the exceptional isolation of GAMA~526784.

Despite its isolation, the gas and star formation properties of GAMA~526784 suggest a complex evolutionary history. The spatial distribution of young clusters, extending over $\sim$5~kpc along a preferential axis, motivates the search for a companion. We hypotehsise that a small galaxy, located approximately 4~arcmin (48.6~kpc) away from GAMA~526784 (Figure~\ref{fig:companion}), may be a promising candidate based on its remarkably similar $g - r = 0.4$ colour to GAMA~526784 and the presence of similar star clusters. 

The putative companion lies within the Subaru Hyper Suprime-Cam (HSC) field of view centred on GAMA~526784, allowing for a detailed analysis of its morphology and physical properties via galaxy fitting, and stellar mass estimation through SED fitting using the five \textit{grizy} HSC filters. The data reduction, as well as the galaxy and SED fitting procedures applied to the companion, are identical to those used for GAMA~526784; we refer the reader to \citetalias{Buzzo_25c} for further methodological details. The analysis yields an effective radius of 0.64~kpc (assuming the same distance as GAMA~526784), a central surface brightness of 23.2~mag~arcsec$^{-2}$ in the $g$-band, and a stellar mass of $\log(M_{\star}/M_{\odot}) = 7.3 \pm 0.2$---roughly an order of magnitude less massive than GAMA~526784. 

GBT observations offer critical constraints on this scenario. While the putative companion falls within the GBT beam, raising the expectation of a second line in the HI spectrum in Figure~\ref{fig:Hi_spectrum} if it were gas-rich, no such line is detected. However, the HI signal at the companion's location would be significantly attenuated by the beam response, reduced to $\sim$40\% of peak sensitivity at a 4.5~arcmin offset. The absence of a second line could indicate one of several possibilities: (1) the companion is gas-poor, (2) its HI line is perfectly blended with that of GAMA~526784, which would require a coincidental velocity alignment, (3) it lies in the background, or (4) the signal is simply too attenuated and faint to detect. The undisturbed HI profile of GAMA~526784 makes blending less likely. The companion’s blue colours suggest recent star formation, making a gas-poor scenario less probable, though it is possible the gas reservoir is small and confined to the immediate vicinity of the star clusters, rather than spread throughout the galaxy. If the companion lies in the background, only deep spectroscopy will be able to resolve this. Still, the presence of extremely similar blue star clusters in both systems, along with the alignment of the companion along the star cluster axis of GAMA~526784, strengthens the case for a physical association, albeit one requiring deeper observations for confirmation.

If physically associated, the system's properties may point to either a high-speed encounter or an ongoing minor merger. The relatively undisturbed morphology of the lower-mass companion is puzzling, as simulations and observations generally predict significant disruption for the smaller galaxy in such interactions. However, this may favour a high-speed flyby scenario, in which the gravitational interaction is brief and insufficient to destroy the companion. The $\sim$100~Myr star formation burst in GAMA~526784 (\citetalias{Buzzo_25c}) implies a relative velocity of $\sim$475~km~s$^{-1}$ to account for the current projected separation of over 4~arcmin, i.e. consistent with a high-speed collision \citep{Lee_24} capable of triggering widespread disk star formation.
Alternatively, a minor merger unrelated to the companion could explain the presence of a two-component stellar population in GAMA526784 (\citetalias{Buzzo_25c}). However, this scenario struggles to account for the formation of massive young clusters, as these clusters are very young, whereas any such merger would have occurred some time ago. Whether gas collapse leading to the formation of massive clusters can be sustained for many millions of years after a minor merger remains an open question, yet to be tested by simulations.

Current data cannot definitively distinguish between these scenarios. The similar optical properties strengthen the case for association, while the GBT non-detection and companion's morphology present challenges. Future observations could resolve this through: (1) deeper HI mapping to detect low-surface-brightness features or confirm the companion's distance, (2) targeting lines that are sensitive to CO-dark molecular gas, and (3) deep spectroscopy of both systems.

GAMA~526784 serves as an intriguing case for studying dwarf galaxy transformations. As shown in Figure~\ref{fig:companion}, the spatial coherence between the galaxy, its hypothesised companion, and star cluster string offers a rare snapshot of a possible interaction, highlighting how even isolated systems may bear traces of their dynamical history.

\section{Conclusions}
\label{sec:conclusions}

Our multi-wavelength study of GAMA~526784 has revealed a remarkable ultra-diffuse galaxy possibly in the midst of transformation, providing new insights into UDG formation and the origins of unusual globular cluster populations in low-mass systems. The galaxy's HI reservoir ($M_{\rm HI} = (4.7\pm0.5)\times10^8$ M$_\odot$) contrasts with the misdetection of CO.
This apparent contradiction may be explained by considering both the prevalence of CO-dark molecular gas expected at its low metallicity ($12+\log(\rm{O/H})=8.3$) and the effects of a possible ongoing interaction with a putative companion approximately 4 arcmin away.

Compelling evidence for this interaction includes the companion's projected separation of 48 kpc, the presence and distribution of shocked gas along the string of young star clusters in GAMA~526784 (\citetalias{Buzzo_25c}), and the alignment of these clusters along the same axis and in the direction of the companion. Their spatial arrangement and ages are consistent with having been triggered during a close encounter approximately 100 Myr ago, which would mean that the galaxies are travelling at $\sim 475$ km s$^{-1}$ with respect to each other, characterising a high-speed encounter. Moreover, the system's two-component structure with an old core surrounded by a younger, star-forming envelope, suggests GAMA~526784 may represent a transitional phase between a gas-rich dwarf and a GC-rich quiescent UDG.

This system challenges simple UDG classification schemes, demonstrating how dwarf-dwarf interactions can produce extended morphologies while simultaneously seeding globular cluster populations. The combination of high velocity dispersion in the gas from the interaction (possibly suppressing global star formation) and local gas compression (enabling star cluster formation) helps explain the apparent paradox of a gas-rich yet inefficiently star-forming galaxy. GAMA~526784 may represent a progenitor phase for some GC-rich UDGs, with its young massive clusters potentially evolving into a globular cluster system over time.

Future investigations should focus on high-resolution HI mapping to reveal tidal structures, lines that are sensitive to CO-dark molecular gas, and deeper spectroscopy of the putative companion to confirm their association. Wider-field searches for similar interacting dwarf pairs could establish how commonly this formation pathway happens. As one of the clearest examples of an isolated UDG caught in the act of transformation, GAMA~526784 provides unique insights into the environmental processes that shape these enigmatic galaxies and their star cluster systems.

\begin{acknowledgements}
AZ acknowledges support from the European Union – NextGeneration EU within PRIN 2022 project n.20229YBSAN - Globular clusters in cosmological simulations and in lensed fields: from their birth to the present epoch and from the INAF Minigrant ‘Clumps at cosmological distance: revealing their formation, nature, and evolution (Ob. Fu. 1.05.23.04.01). MH acknowledges financial support from the Deutsche Forschungsgemeinschaft (DFG, German Research Foundation) under Germany’s Excellence Strategy – EXC-2094 – 390783311. KS acknowledges support from the Natural Sciences and Engineering Research Council of Canada. 
\end{acknowledgements}

\bibliographystyle{aa} 
\bibliography{bibli.bib}
\begin{appendix}

\section{Diffuse body (low-resolution data) CO map}
\label{sec:appendix_diffuse}
\begin{figure}
    \centering
    \includegraphics[width=\columnwidth, trim=0 0 10cm 0, clip]{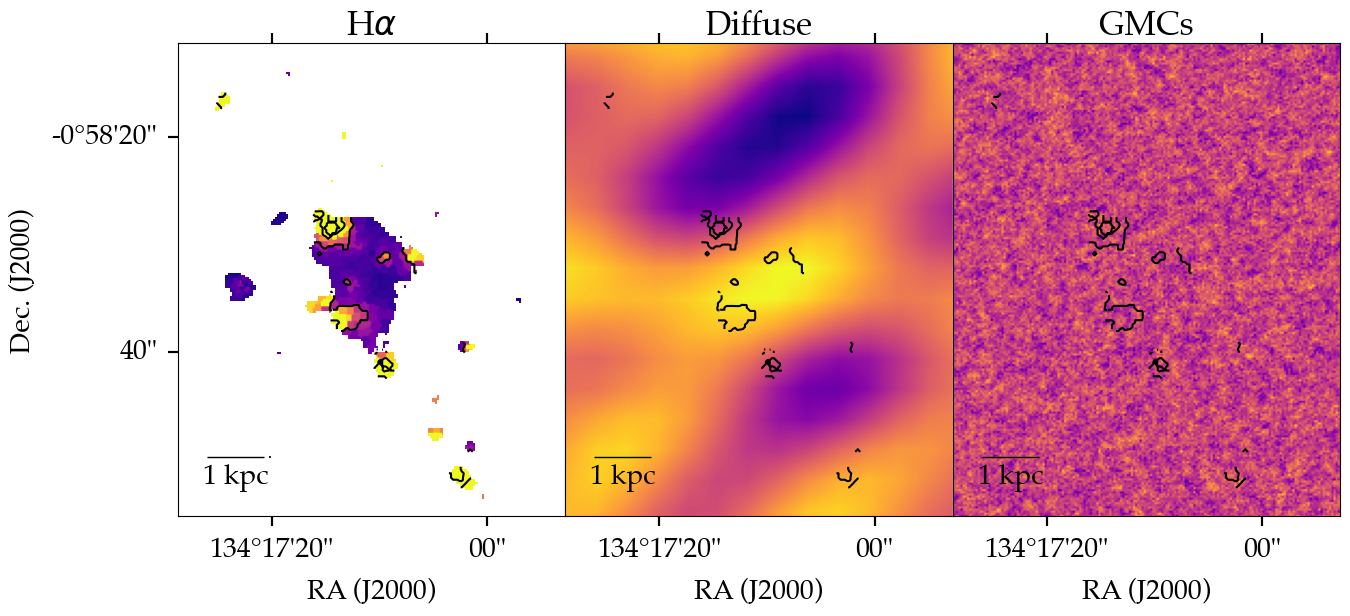}
    \caption{Comparison of H$\alpha$ map, highlighting where the molecular gas would be expected in the diffuse body if it was co-spatial with the ionised gas. The right panel shows the averaged CO map of the low-resolution ALMA data, focused on the diffuse body. In both panels, H$\alpha$ contours are shown to facilitate visualization.}
    \label{fig:diffuse_GMCs}
\end{figure}

In this appendix, we present the averaged CO map of the low-resolution data targetting the diffuse body to show the lack of any clear CO emission throughout the galaxy, shown in Figure~\ref{fig:diffuse_GMCs}.

\section{Table with star cluster properties}
In this appendix, we show the physical properties of GAMA~526784 and its star clusters measured in \citetalias{Buzzo_25c} and in this work.

\begin{table*}
    \centering
    \caption{Physical properties of GAMA~526784 and its star clusters. Columns show: (1) Source ID from \citetalias{Buzzo_25c}; (2) $5\sigma$ CO(1-0) flux upper limits; (3) Star formation rates from H$\alpha$ measurements; (4) Gas-phase metallicities; (5) Stellar masses; (6) Molecular gas mass upper limits; (7) Neutral gas mass (for the main galaxy only). }
    \scalebox{0.9}{
    \begin{tabular}{ccccccc} \hline
        Source & CO(1-0)$_{5\sigma}$ ($\mu$Jy) & SFR ($10^{-5}$ M$_\odot$ yr$^{-1}$) & $12 + \log(\rm[O/H])$ & $\log(M_{\star}/M_{\odot})$ & $\log(M_{H_2}/M_{\odot})_{\alpha_{\rm CO, met}}$ & $\log(M_{HI}/M_{\odot})$ \\ \hline
        GAMA~526784 & $\leq$5.80 & 100 & 8.30 & $8.24^{+0.12}_{-0.23}$ & $\leq$$7.60$ & $8.70 \pm 0.90$ \\
        Globular Cluster 1 & $\leq$1.88 & -- & -- & 5.64 & -- & -- \\
        Star Cluster 2 & $\leq$0.87 & 7.49 & 8.09 & 5.19 & $\leq$7.09 & -- \\
        Globular Cluster 3 & $\leq$0.61 & -- & -- & 5.90 & -- & -- \\
        Globular Cluster 4 & $\leq$0.82 & -- & -- & 5.93 & -- & -- \\
        Globular Cluster 5 & $\leq$1.14 & -- & -- & 5.89 & -- & -- \\
        Star Cluster 6 & $\leq$0.57 & 0.73 & 8.24 & 5.75 & $\leq$6.59 & -- \\
        Star Cluster 7 & $\leq$1.03 & 1.48 & 8.39 & 4.88 & $\leq$6.64 & -- \\
        Star Cluster 8 & $\leq$0.85 & 34.00 & 8.20 & 5.33 & $\leq$6.93 & -- \\
        Star Cluster 9 & $\leq$0.90 & 8.55 & 8.34 & 5.06 & $\leq$6.71 & -- \\
        Globular Cluster 10 & $\leq$0.57 & -- & -- & 5.77 & -- & -- \\
        Star Cluster 11 & $\leq$1.03 & 1.45 & 8.39 & 4.50 & $\leq$6.66 & -- \\
        Star Cluster 12 & $\leq$0.84 & 33.00 & 8.23 & 5.50 & $\leq$6.88 & -- \\
        Star Cluster 13 & $\leq$0.59 & 0.86 & 8.36 & 5.49 & $\leq$6.41 & -- \\
        Star Cluster 14 & $\leq$0.66 & 3.08 & 8.32 & 4.60 & $\leq$6.61 & -- \\
        Star Cluster 15 & $\leq$0.71 & 4.27 & 8.30 & 4.93 & $\leq$6.66 & -- \\
        Star Cluster 16 & $\leq$0.60 & 0.98 & 8.40 & 5.73 & $\leq$6.34 & -- \\
        Star Cluster 17 & $\leq$0.60 & 0.71 & 8.30 & 4.17 & $\leq$6.56 & -- \\
        Star Cluster 18 & $\leq$0.58 & 2.13 & 8.39 & 5.69 & $\leq$6.37 & -- \\
        Star Cluster 19 & $\leq$0.61 & 19.00 & 8.24 & 5.52 & $\leq$6.70 & -- \\
        Star Cluster 20 & $\leq$0.62 & 18.00 & 8.27 & 5.47 & $\leq$6.66 & -- \\
        Globular Cluster 21 & $\leq$0.62 & -- & -- & 5.69 & -- & -- \\
        Star Cluster 22 & $\leq$0.64 & 1.43 & 8.41 & 4.27 & $\leq$6.43 & -- \\
        Star Cluster 23 & $\leq$0.60 & 17.00 & 8.17 & 5.05 & $\leq$6.83 & -- \\
        Star Cluster 24 & $\leq$0.58 & 7.74 & 8.32 & 4.55 & $\leq$6.57 & -- \\ 
        Star Cluster 25 & $\leq$0.58 & 5.71 & 8.37 & 4.58 & $\leq$6.48 & -- \\
        Star Cluster 26 & $\leq$0.61 & 2.62 & 8.25 & 4.91 & $\leq$6.66 & -- \\
        Star Cluster 27 & $\leq$0.92 & 9.37 & 8.17 & 4.65 & $\leq$7.01 & -- \\
        Globular Cluster 28 & $\leq$1.01 & -- & -- & 5.50 & -- & -- \\
        Globular Cluster 29 & $\leq$0.71 & -- & -- & 5.83 & --  & -- \\ \hline
    \end{tabular}}
    \label{tab:sfr_metallicity}
\end{table*}

\end{appendix}
\end{document}